\documentclass[twocolumn,english,aps,prl,superscriptaddress]{revtex4-2}
\usepackage[T1]{fontenc}
\usepackage{amsmath,graphicx,amssymb,epsfig,babel,dsfont}
\usepackage{mathrsfs}
\usepackage{xcolor}
\usepackage[normalem]{ulem}

\renewcommand{\Re}{{\rm Re}}
\renewcommand{\Im}{{\rm Im}}

\newcommand{\rd}{{d}}
\newcommand{\kB}{k_{B}}

\newcommand{\ri}{{i}}

\newcommand{\re}{{e}}

\begin{document}

\title{Feynman paradox induced by vacuum and thermal fluctuations}

\author{Svend-Age Biehs}
\email{s.age.biehs@uni-oldenburg.de}
\affiliation{Institut f\"{u}r Physik, Carl von Ossietzky Universit\"{a}t, 26111, Oldenburg, Germany}

\author{Ivan Latella}
\email{ilatella@ub.edu}
\affiliation{Departament de F\'{i}sica de la Mat\`{e}ria Condensada, Universitat de Barcelona, Mart\'{i} i Franqu\`{e}s 1, 08028 Barcelona, Spain}
\affiliation{Institut de Nanociència i Nanotecnologia de la Universitat de Barcelona (IN2UB), Diagonal 645, 08028 Barcelona, Spain}

\begin{abstract}
A charged particle initially at rest in an external magnetic field starts to rotate when the magnetic field is switched off. This is a variant of the Feynman disc paradox, where the conservation of angular momentum is seemingly violated. The paradox is understood by realizing that angular momentum is initially stored in the electromagnetic field and is transferred to the particle when the magnetic field is removed.
In a classical description, no rotation occurs if the particle is uncharged, as the initial angular momentum is zero in this case.
We show that electromagnetic fluctuations in thermal equilibrium can induce a quantum analog of the Feynman paradox, where a nonreciprocal particle without charge starts to rotate when the source of nonreciprocity is removed.
This paradox is due to persistent energy fluxes arising in nonreciprocal systems at equilibrium, leading to angular momentum stored in the electromagnetic field.
We demonstrate that the contribution of vacuum fluctuations to persistent energy fluxes dominate over thermal fluctuations at finite temperature, so vacuum fluctuations dominate the equilibrium angular momentum as well. Observation of the induced motion would thus provide a means of detecting persistent energy fluxes and offer further evidence for the physical reality of vacuum fluctuations.
\end{abstract}

\maketitle

Fluctuations of the electromagnetic field are fundamental carriers of energy and momentum that drive a wide range of phenomena within light-matter interactions~\cite{Rytov,Casimir,Casimir_Polder,Lifshitz,Polder}. In systems composed of nonreciprocal materials~\cite{Caloz,Yang_2024}, such as magneto-optical (MO) media~\cite{Armelles_2013,Moncada} or Weyl semimetals (WSMs)~\cite{Kotov_2018,Guo_elight}, these fluctuations give rise to distinctive effects including directional radiative heat transfer~\cite{PBAHall,Latella_2017,OTTReview,Ott_2020,He_2020,Zhou_2020,Dong_2021,Cuevas,Zhang_2022,Biehs_2023,Latella_2025} and fluctuation-induced torque~\cite{Pang_2019,Guo_2021,Khandekar_2021,Gao_2021,Strekha_2022,Milton_2023,Strekha_2024} in nonequilibrium conditions. Even in global thermal equilibrium, nonreciprocal media exhibit remarkable behavior through the emergence of persistent energy fluxes~\cite{ZhuFan,Silveirinha,ZhuEtAl2018,OTTcircular,Khandekar_2019,Biehs_2025}, consisting of circulating flows of energy with no net dissipation, that are entirely absent in reciprocal systems. 

In nonreciprocal electromagnetic systems at equilibrium, persistent energy fluxes are characterized by a nonvanishing mean Poynting vector (PV)~\cite{ZhuFan}. Apparent energy fluxes can also arise in classical electrostatic systems where nonzero PVs can exist. A notable example is a paradox introduced by Feynman~\cite{Feynman}, considering a charged disc that is initially at rest under the influence of an external magnetic field. Under these conditions, Feynman claimed that the disc would begin to rotate when the magnetic field was switched off, seemingly violating the conservation of angular momentum (AM). The paradox is resolved by noting that AM is initially stored in the electromagnetic field and subsequently transferred to the disc when the external field is removed~\cite{Feynman,Aguirregabiria,Padmanabhan}. This stored AM originates from a nonzero PV, as noted above, produced by the electric field of the disc's charge and the external magnetic field. A similar effect would be observed if the disc were replaced by a charged metallic nanoparticle.

In this Letter, we show that both vacuum and thermal fluctuations of the electromagnetic field contribute to the PV in global thermal equilibrium, resulting in electromagnetic AM around a neutral but non-reciprocal particle. This AM has no counterpart in the purely classical description, yet, as in the classical case, it can induce motion once the nonreciprocity is removed. The phenomenon can thus be regarded as a quantum analog of the Feynman paradox in the absence of a net charge.
Our findings also reveal that the contribution of vacuum fluctuations dominates the equilibrium PV and associated AM at finite temperature, implying that persistent energy fluxes persist even in the zero temperature limit.

In the classical description [see Figs.~1(a) and (b)] of a metallic nanoparticle with volume $V$ and charge $q$ producing an electric field $\mathbf{E}_q$, the PV at a point $\mathbf{r}$ outside the particle is given by $\mathbf{S} =\mu_0^{-1} \mathbf{E}_{q}\times\mathbf{B}_e$, where $\mu_0$ is the vacuum permeability and $\mathbf{B}_e$ is the external magnetic field. While $\mathbf{S}$ represents an energy flux, the quantity $\mathbf{r}\times \mathbf{S}/c^{2}$ accounts for the AM density of the electromagnetic field~\cite{Jackson}, $c$ being the speed of light. Thus, the AM transferred to the particle in Feynman's argument is $\mathbf{J}=\int_{\mathbf{r}\notin V} \mathbf{r}\times (\mathbf{S}/c^2)\, d^3r$. Since $\mathbf{E}_q=\mathbf{0}$ everywhere if $q=0$, for a neutral particle one has $\mathbf{J}=\mathbf{0}$ regardless of the value of $\mathbf{B}_e$.

\begin{figure}
    \centering
	\includegraphics[width=\columnwidth]{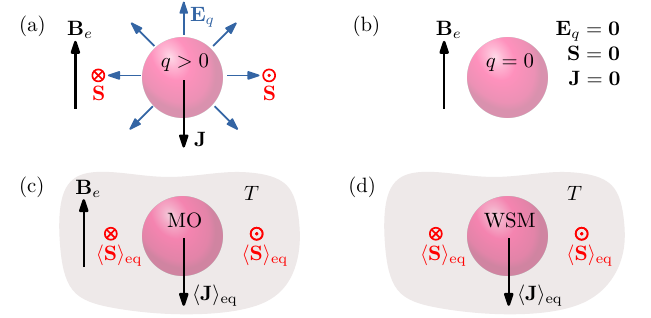}
	\caption{AM in the Feynman paradox: (a) Classical setup for a nanoparticle with charge $q$, where the Poynting vector $\mathbf{S}$ and the electromagnetic angular momentum $\mathbf{J}$ result from the electric field $\mathbf{E}_q$ produced by the charge and the external magnetic field $\mathbf{B}_e$. (b) In the classical setup, both $\mathbf{S}$ and $\mathbf{J}$ vanish if the particle is uncharged because $\mathbf{E}_q=\mathbf{0}$. (c) Considering vacuum and thermal fluctuations of the electromagnetic field at temperature $T$, an uncharged magneto-optical (MO) particle in presence of an external magnetic field leads to a nonzero mean Poynting vector $\langle \mathbf{S} \rangle_{\rm eq}$, so the electromagnetic angular momentum $\langle \mathbf{J} \rangle_{\rm eq}$ is nonzero as well. (d) For intrinsically nonreciprocal media like a Weyl semimetal (WSM), nonvanishing $\langle \mathbf{S} \rangle_{\rm eq}$ and $\langle \mathbf{J} \rangle_{\rm eq}$ also exists without any external magnetic field.
		\label{Fig:Sketch} }
\end{figure}

Now, instead of a charged object, we consider a globally neutral, nonmagnetic nanoparticle at rest made of a nonreciprocal material and thermalized with the environment at temperature $T$  [see Figs.~1(c) and (d)]. The nanoparticle here is modeled as an electric dipole occupying a small volume $V$. As we discuss below, the use of nonreciprocal materials can lead to electromagnetic energy fluxes that persist in thermal equilibrium~\cite{ZhuFan,Silveirinha,ZhuEtAl2018,OTTcircular,Khandekar_2019,Biehs_2025}, manifested as a nonzero mean PV $\langle \mathbf{S} (\mathbf{r})\rangle_{\rm eq}$. Taking into account the quantum nature of electromagnetic radiation, $\langle \mathbf{S} (\mathbf{r})\rangle_{\rm eq}$ is understood as an ensemble average over fluctuations of the electric and magnetic fields. Since the momentum density inside the particle can be neglected in the dipole approximation (see End Matter), the corresponding electromagnetic AM is
\begin{equation}
	\langle \mathbf{J} \rangle_{\rm eq} = \frac{1}{c^2}\int_{\mathbf{r}\notin V} \rd^3 r\, \mathbf{r} \times \langle \mathbf{S}(\mathbf{r}) \rangle_{\rm eq}.
    \label{Eq:Jeq}
\end{equation}
When the system is reciprocal, $\langle \mathbf{S}(\mathbf{r}) \rangle_{\rm eq}$ is zero and $\langle \mathbf{J} \rangle_{\rm eq}$ vanishes.
Furthermore, a nonreciprocal particle in thermal equilibrium experiences no net torque~\cite{Pang_2019,Guo_2021,Gelbwaser-Klimovsky_2022,Manjavacas_2010_1,Manjavacas_2010_2}, implying that the system's total AM remains conserved.
In MO media such as doped semiconductors, reciprocity can be broken by the presence of an external magnetic field $\mathbf{B}_e$ [Fig.~\ref{Fig:Sketch}(c)]. Hence, the removal of the external field must induce a rotation of the particle, provided it is initially at rest, ensuring the conservation of AM. For nonreciprocal WSM particles, the induced rotation is \textit{anomalous} because nonzero $\langle \mathbf{S} \rangle_{\rm eq}$ and $\langle \mathbf{J} \rangle_{\rm eq}$ are obtained without external magnetic field [Fig.~\ref{Fig:Sketch}(d)].
Driving the WSM through a phase transition, for example, via a temperature change, would make the particle reciprocal, causing the motion of the particle to preserve the total AM.

When fluctuations of the electromagnetic field are considered in equilibrium at temperature $T$, the mean PV can be expressed in components by
\begin{equation}
	\langle S_\nu(\mathbf{r}) \rangle_{\rm eq} = \epsilon_{\nu \alpha \beta} \langle E_\alpha (\mathbf{r}, t) H_\beta(\mathbf{r}, t) \rangle_{\rm eq}^{\rm sym}, 
	\label{Eq:Pontingvector_components}
\end{equation}
where we use the summation convention for repeated indices, $ \epsilon_{\nu \alpha \beta}$  being the Levi-Civita symbol. Here, $E_\alpha (\mathbf{r}, t)$ and $H_\beta(\mathbf{r}, t)$ are the components of the electric and magnetic fields at a point $\mathbf{r}$ and time $t$, which include the fluctuating fields emitted by the particle and background environmental fields.
Within the fluctuation electrodynamics approach, the symmetrically ordered correlation function on the right-hand side of Eq.~(\ref{Eq:Pontingvector_components}) is obtained from the fluctuation-dissipation theorem~\cite{Callen-Welton} using linear response theory~\cite{Eckhardt,GSA}. For a single spherical nanoparticle in vacuum at position $\mathbf{r}_p$ with polarizability tensor $\underline{\alpha}$, this correlation function takes the form~\cite{Biehs_2025}
\begin{equation}
		\langle E_\alpha (\mathbf{r}, t) H_\beta(\mathbf{r}, t) \rangle_{\rm eq}^{\rm sym} = \int_0^\infty \frac{\rd \omega}{2 \pi} \frac{2k_0^3}{\varepsilon_0 c} \Theta(\omega,T) \Im \mathds{Y}_{\alpha\beta}(\mathbf{r};\omega),
		\label{Eq:correlation_function}
\end{equation}
where $\varepsilon_0$ is the vacuum permittivity and
\begin{equation}
   \Theta(\omega,T) = \frac{\hbar \omega}{\re^{\hbar\omega/\kB T} - 1} + \frac{\hbar\omega}{2}=\frac{\hbar\omega}{2}\coth\left(\frac{\hbar\omega}{2k_BT}\right)
\end{equation}
is the mean energy of a harmonic oscillator at frequency $\omega$. Here, $k_0=\omega/c$ is the vacuum wavenumber, with $\hbar$ and $\kB$ being the reduced Planck constant and Boltzmann constant, respectively. 
In Eq.~(\ref{Eq:correlation_function}), we have introduced the components of the matrix (see End Matter)
\begin{equation}
	\mathds{Y}(\mathbf{r};\omega) = 
	\mathds{G}^{\rm EE, vac}(\mathbf{r}, \mathbf{r}_p) \bigl(\underline{\alpha} - {\underline{\alpha}}^t \bigr) \mathds{G}^{\rm HE, vac}( \mathbf{r},\mathbf{r}_p ) 
	\label{Eq:matrix_Y}
\end{equation}
for points $\mathbf{r}\notin V$, which is defined in terms of the electric-electric and magnetic-electric dyadic Green's functions in vacuum $\mathds{G}^{\rm EE, vac}$ and $\mathds{G}^{\rm HE, vac}$; the superscript $t$ in the polarizability denotes the transpose. It is clear that expression~(\ref{Eq:matrix_Y}) is nonzero if the polarizability is nonreciprocal, meaning that $\underline{\alpha} \neq {\underline{\alpha}}^t$, which is inherited from the permittivity tensor when the time-reversal symmetry is broken.
Furthermore, assuming that the particle is in the origin at $\mathbf{r}_p = \mathbf{0}$, the vacuum Green's functions in spherical coordinates with basis $\{\mathbf{e}_r,\mathbf{e}_\theta,\mathbf{e}_\varphi\}$ read~\cite{OTTcircular}
\begin{align}
	\mathds{G}^{\rm EE, vac}(\mathbf{r}, \mathbf{r}_p) &= \frac{\re^{\ri k_0 r}}{4 \pi r} \bigl[ a \mathds{1} + b \mathbf{e}_r\otimes\mathbf{e}_r \bigr], \\
	\mathds{G}^{\rm HE, vac}(\mathbf{r}, \mathbf{r}_p) &= \sqrt{\frac{\varepsilon_0}{\mu_0}} \frac{\re^{\ri k_0 r}}{4 \pi r} l \bigl[ \mathbf{e}_\varphi \otimes \mathbf{e}_\theta - \mathbf{e}_\theta\otimes\mathbf{e}_\varphi \bigr],
\end{align}
where $r=|\mathbf{r}|$ and
\begin{equation}
	a = 1 + \frac{\ri k_0 r - 1}{k_0^2 r^2},\quad b = \frac{3 - 3 \ri k_0 r - k_0^2 r^2}{k_0^2 r^2},\quad l = 1 + \frac{\ri}{k_0 r}.
\end{equation}
Hence, using Eqs.~(\ref{Eq:correlation_function}) and (\ref{Eq:matrix_Y}) in Eq.~(\ref{Eq:Pontingvector_components}) with the polarizability in the spherical basis, we obtain
\begin{equation}
\begin{split}
	\langle \mathbf{S} (\mathbf{r})\rangle_{\rm eq} &=  \int_0^\infty \frac{\rd \omega}{2 \pi} \, \Theta(\omega,T) k_0^3  
	2 \Im \biggl\{ \frac{\re^{2 \ri k_0 r}}{(4 \pi r)^2} l (a + b)\\
	&\times \bigl[  (\alpha_{r \theta} - \alpha_{\theta r}) \mathbf{e}_\theta + (\alpha_{r \varphi} - \alpha_{\varphi r}) \mathbf{e}_\varphi \bigr] \biggr\} 
\end{split}
\label{Eq:MeanPVparticle}
\end{equation}
for $r>R_p$, $R_p$ being the radius of the particle. This PV has no radial component; therefore, the particle neither emits nor absorbs energy in thermal equilibrium, despite the persistent energy flux around it.

For the specific case of MO materials or WSMs where the magnetic field or the Weyl node vector lies in the $z$-direction, the polarizability tensor in Cartesian coordinates has the form~\cite{Palik,Moncada,Kotov_2016,Kotov_2018,Guo_elight}
\begin{equation}
   \underline{\alpha} = \begin{pmatrix} \alpha_{11} & \alpha_{12} & 0 \\ -\alpha_{12} & \alpha_{11} & 0 \\ 0 & 0 & \alpha_{33}	\end{pmatrix}.
   \label{Eq:polarizabilty_single_particle}
\end{equation}
Using this, Eq.~(\ref{Eq:MeanPVparticle}) simplifies to $\langle \mathbf{S} (\mathbf{r})\rangle_{\rm eq} = S_\varphi(r,\theta) \mathbf{e}_\varphi$ with the azimuthal component given by
\begin{equation}
S_\varphi(r,\theta) = \Im \int_0^\infty \rd \omega\, s(\omega)
\label{Eq:Sr}
\end{equation}
and
\begin{equation}
s(\omega)=\frac{\hbar \sin\theta}{8 \pi^3 r^5} \coth\biggl(\frac{\hbar\omega}{2k_BT}\biggr) 
	     \ri  \omega \re^{2 \ri \frac{\omega}{c} r}\biggl( 1- \frac{\ri \omega r}{c}\biggr)^2 \alpha_{12}(\omega). 
\label{Eq:small_s} 
\end{equation}
We see that $\langle \mathbf{S} (\mathbf{r})\rangle_{\rm eq}$ is purely circular and energy flows around the axis parallel to the magnetic field~\cite{OTTcircular}. Notice that $\alpha_{12}$ can be nonzero in absence of dissipation within the nanoparticle, so a persistent energy flux can exist even if the material is lossless. The flux $\langle\mathbf{S}(\mathbf{r})\rangle_{\rm eq}$ vanishes when the material is reciprocal, for which $\alpha_{12}=0$.

For a single nanoparticle with nonreciprocal polarizability tensor $\underline{\alpha}$, we show below that the contribution of vacuum fluctuations to the equilibrium mean PV is finite, so the persistent energy flux also persists at zero temperature. 
To compute $S_\varphi$ in Eq.~(\ref{Eq:Sr}), we evaluate the integral $\int_\mathcal{C} \rd \omega\, s(\omega)$ along a contour $\mathcal{C}$ in the first quadrant of the complex $\omega$-plane, following an approach analogous to that used in calculating fluctuation-induced forces between neutral bodies~\cite{Lifshitz}. Assuming that $\alpha_{12}$ is analytic in the upper half of the plane, we only need to consider the poles of $s(\omega)$ arising from $\coth\left(\frac{\hbar\omega}{2k_BT}\right)$. These poles are located on the imaginary axis at frequencies $\omega=\ri \xi_n$ and are avoided in the contour $\mathcal{C}$ with semicircles, where $\xi_n=n 2\pi k_B T/\hbar$ are the Matsubara frequencies, $n$ being an integer.
In addition, the integral of $s(\omega)$ vanishes at infinite complex frequencies by assuming that $\omega^3\alpha_{12}(\omega)\to 0$ as $|\omega|\to\infty$ (see End Matter), and the integral on the imaginary axis is a real quantity because $\alpha_{12}(\omega)$ is real for purely imaginary $\omega$. Accordingly, when taking the imaginary part, we obtain
\begin{equation}
\Im \int_0^\infty \rd \omega\, s(\omega)=\Im\left(\ri\pi \sum_{n=1}^\infty\mathrm{Res}(s,\ri\xi_n)\right),
\end{equation}
where $\mathrm{Res}(s,\ri\xi_n)$ is the residue of $s(\omega)$ at $\omega=\ri\xi_n$. Notice that the residue at the origin is zero, as we assumed that $\alpha_{12}(0)$ is bounded.
The remaining residues can be evaluated, in such a way that introducing the characteristic thermal length $\ell_T  =\hbar c/2\pi k_B T$ (so $\xi_n=n c/\ell_T$)
and the polarizability per unit volume
\begin{equation}
\eta_n  =\eta(\ri\xi_n) ,\qquad
\eta(\omega) =\frac{3\alpha_{12}(\omega)}{4\pi R_p^3} ,
\end{equation}
the azimuthal component of $\langle \mathbf{S} (\mathbf{r})\rangle_{\rm eq}$ becomes
\begin{equation}
S_\varphi(r,\theta) = -\frac{ \hbar c^2 R_p^3 \sin\theta }{ 6 \pi^2\ell_T^2 r^5} 
          \sum_{n=1}^\infty n \re^{-2 n r/\ell_T}
	   \left( 1+\frac{n r}{\ell_T} \right)^2 \eta_n  .
	   \label{Eq:Sr_summation_2}
\end{equation}
Expression~(\ref{Eq:Sr_summation_2}) is exact within the dipole approximation and accounts for both vacuum and thermal fluctuations of the electromagnetic field. 

The contribution of vacuum fluctuations can be explicitly obtained by taking the limit $T\to0$ in Eq.~(\ref{Eq:Sr_summation_2}). In this limit, the distance between poles approaches zero and the summation over $n$ can be replaced by an integration over $\xi$ such that $\sum_{n}\to \frac{\hbar}{2\pi k_BT}\int_0^\infty\rd\xi $~\cite{Lifshitz}.
Denoting $S_0=\lim_{T\to0}S_\varphi$, the azimuthal component of the persistent energy flux at zero temperature reads 
\begin{equation}
S_0(r,\theta) = -\frac{\hbar R_p^3\sin\theta}{6 \pi^2 r^5}\int_0^\infty \rd\xi\,  \xi \re^{-2 \frac{\xi}{c} r}
	   \left( 1+\frac{ \xi r}{c}\right)^2 \eta(\ri\xi).  
\label{Eq:S0_imaginary}	   
\end{equation}
At finite temperature, we then identify the thermal contribution to the persistent flux $\langle \mathbf{S} (\mathbf{r})\rangle_T=S_T(r,\theta) \mathbf{e}_\varphi$, such that $S_T(r,\theta)=S_\varphi(r,\theta)-S_0(r,\theta)$, which can be computed by means of Eqs.~(\ref{Eq:Sr_summation_2}) and (\ref{Eq:S0_imaginary}).

\begin{figure}
    \centering
	\includegraphics[width=\columnwidth]{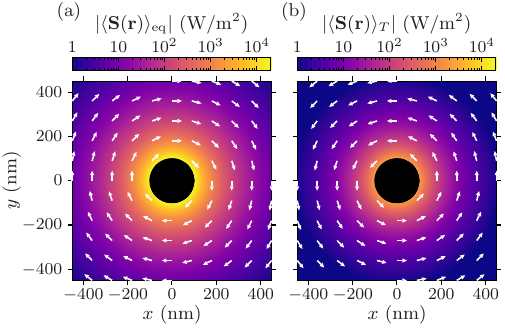}
	\caption{Persistent energy flux around a nonreciprocal InSb nanoparticle of radius $R_p=100\,$nm in equilibrium at $T=300\,$K with a magnetic field $B_e=1\,$T applied along the positive $z$-axis. In (a) we show $\langle \mathbf{S} \rangle_{\rm eq} = S_\varphi \mathbf{e}_\varphi$ with $S_\varphi$ obtained via Eq.~(\ref{Eq:Sr_summation_2}), while in (b) we represent only the thermal contribution $\langle \mathbf{S} \rangle_T=S_T \mathbf{e}_\varphi$ with $S_T=S_\varphi-S_0$ computed by means of Eqs.~(\ref{Eq:Sr_summation_2}) and (\ref{Eq:S0_imaginary}). Here the polar angle is $\theta=\pi/2$.
		\label{Fig:Seq} }
\end{figure}

In End Matter, we describe the polarizability tensor of MO doped semiconductors such as InSb, subject to an external magnetic field of magnitude $B_e$ oriented along the $z$-axis~\cite{Palik,Moncada,Carminati,Albaladejo}. Using this model, in Fig.~\ref{Fig:Seq} we show $\langle \mathbf{S} (\mathbf{r})\rangle_{\rm eq}$ and the corresponding thermal contribution $\langle \mathbf{S} (\mathbf{r})\rangle_T$ around an InSb particle of radius $R_p=100\,$nm at $T=300\,$K. As shown in the figure, the thermal contribution circulates counterclockwise, whereas the total energy flux $\langle \mathbf{S} (\mathbf{r})\rangle_{\rm eq}$ is dominated by vacuum fluctuations which results in a net clockwise circulation.

\begin{figure}
    \centering
	\includegraphics[width=\columnwidth]{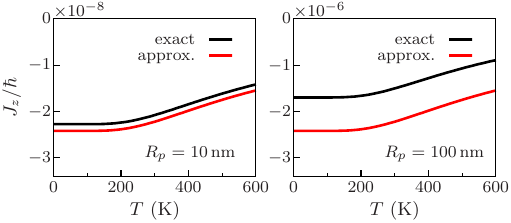}
	\caption{AM of the electromagnetic field around a non-reciprocal InSb nanoparticle in equilibrium at temperature $T$. A magnetic field $B_e=1\,$T is applied along the positive $z$-axis. Here, $J_z$ is computed using the exact expression in Eq.~(\ref{Eq:Jz}) and the approximation in Eq.~(\ref{Eq:Jz_approximation}).
		\label{Fig:Jz} }
\end{figure}

The preceding discussion allows us to conclude that the persistent energy flux $\langle \mathbf{S} (\mathbf{r})\rangle_{\rm eq}$ can exist even in the absence of thermal fluctuations, and therefore is not necessarily a flow of heat. Rather, it can be regarded as an apparent energy flux that contributes to the electromagnetic AM density in Eq.~(\ref{Eq:Jeq}), as in the classical Feynman paradox.
In addition, Eq.~(\ref{Eq:Jeq}) can be exactly integrated using Eq.~(\ref{Eq:Sr_summation_2}), so $\langle \mathbf{J} \rangle_{\rm eq}=J_z \mathbf{e}_z$ with
\begin{equation}
		J_z = -  \hbar\frac{2 x^2}{9 \pi }  \sum_{n=1}^\infty n\re^{-2 n x } \left( 2 +  n x  \right) \eta_n, 
\label{Eq:Jz}
\end{equation}
where $x = R_p/\ell_T$. This AM stems from the quantum nature of the electromagnetic field and has no classical analogue.
Furthermore, in End Matter we show that for a MO semiconductor, $\eta(\ri\xi)$ can be approximated as
\begin{equation}
\eta(\ri\xi)\approx\frac{9\varepsilon_\infty  \omega_p^2 \omega_c(B_e) \xi}{[(\varepsilon_\infty+2)\xi^2+\varepsilon_\infty\omega_p^2]^2},
\label{Eq:eta_approx}
\end{equation}
with $\varepsilon_\infty$ the high-frequency dielectric constant, $\omega_p$ the plasma frequency of carriers of effective mass $m^*$, and $\omega_c(B_e)=eB_e/m^*$ the cyclotron frequency, where $e$ is the elementary charge. Equation~(\ref{Eq:eta_approx}) allows us to evaluate $\eta_n  =\eta(\ri\xi_n)$ and compute $J_z$ for small particle radii $R_p\ll \ell_T$. By introducing the material's plasma temperature $T_p= \frac{\hbar \omega_p}{ 2 k_B }  \frac{\sqrt{\varepsilon_\infty}}{\sqrt{\varepsilon_\infty+2}}$ and taking the asymptotic limit $x = R_p/\ell_T\to 0$ in Eq.~(\ref{Eq:Jz}), we obtain
\begin{equation}
	J_z \approx - \hbar \frac{R_p^2 \omega_p }{c^2} \frac{\sqrt{\varepsilon_\infty}   f(T)}{\sqrt{(\varepsilon_\infty+2)^3}} \omega_c(B_e)
\label{Eq:Jz_approximation}           
\end{equation}
with
\begin{equation}
	f(T)= \frac{4}{\pi^2}  \frac{T_p}{T} \sum_{n=1}^\infty \frac{ n^2}{\big[n^2  +(T_p/\pi T)^2\big]^2}.
\end{equation}
The series above can be summed exactly, yielding $f(T)=\coth\left(\frac{ T_p}{T}\right)- \frac{ T_p}{T} \,\mathrm{csch}^2\left(\frac{T_p}{T}\right)$, which approaches unity as $T\to 0$ and vanishes in the infinite temperature limit.
For our InSb model, we have $T_p=669\,$K and $f(T)=0.92$ at $T=300\,$K, while $\ell_T=1.2\,\mu$m at the same temperature. Thus, it turns out that a nanoparticle with $R_p=10\,$nm at room temperature and a field $B_e=1\,$T along the positive $z$-axis produces an electromagnetic AM $J_z/\hbar\approx-2\times10^{-8}$, much smaller than the spin of a single electron. A comparison between Eq.~(\ref{Eq:Jz_approximation}) and the exact result in Eq.~(\ref{Eq:Jz}) is shown in Fig.~\ref{Fig:Jz} for particles of radius $R_p=10\,$nm and $R_p=100\,$nm and an applied field of magnitude $B_e=1\,$T. In this figure, we observe that $J_z$ increases as the temperature increases, due to the effect of thermal fluctuations that oppose the dominant contribution from vacuum fluctuations.

Now suppose that the particle is initially at rest under the action of the applied magnetic field. When the magnetic field is switched off, the initial AM stored in the electromagnetic field is transferred to the particle, which then rotates with angular velocity $\Omega=|J_z|/I$, where $I=2 M_pR_p^2/5$ is the particle's moment of inertia and $M_p$ its mass. In the approximation~(\ref{Eq:Jz_approximation}) for $R_p\ll \ell_T$, the rotation frequency can be computed as
\begin{equation}
	\Omega \approx \frac{5}{2}\frac{\hbar \omega_p}{M_p c^2} \frac{\sqrt{\varepsilon_\infty} f(T) }{ \sqrt{(\varepsilon_\infty+2)^3}}\omega_c(B_e)  .
\label{Eq:Omega_approx}
\end{equation}
Hence, with a mass density $\rho=5.8\times10^{3}\,$kg/m$^3$ for InSb and for a small particle of radius $R_p=10\,$nm, a frequency $\Omega\approx 2.4\times 10^{-6}\,$rad/s is achieved at room temperature for an initial field $B_e=1\,$T (before it is turned off). As seen from Eq.~(\ref{Eq:Omega_approx}), this small value of $\Omega$ can be attributed the relatively low energy of electromagnetic fluctuations $\hbar\omega_p$ compared to the particle’s rest energy $M_p c^2$.

In conclusion, we have demonstrated that both vacuum and thermal fluctuations contribute to persistent energy fluxes around a nonreciprocal nanoparticle in thermal equilibrium. These fluxes exist even in the absence of dissipation within the particle and persist down to zero temperature, indicating that they do not represent a genuine flow of heat. Instead, persistent fluxes store a finite equilibrium AM, which can be transferred to the particle when the source of nonreciprocity is removed; for instance, by switching off an external magnetic field in MO materials or inducing a temperature-driven phase transition in WSMs. A similar effect is also expected for a MO thin film. When deposited on a microcantilever, the associated AM transfer could be detected using an approach analogous to Einstein-de Haas effect measurements~\cite{Wallis}. Periodic reversal of the magnetic field would drive cantilever oscillations, providing a direct mechanical signature of the AM stored in the electromagnetic field and a pathway to experimentally detect persistent energy fluxes in equilibrium. These findings establish a previously unexplored connection between fluctuation-induced momentum and the motion of isolated bodies, grounded in fundamental conservation laws.

%
%

\begin{acknowledgments}
S.-A.\ Biehs gratefully acknowledges financial support from the Niedersächsische Ministerium für Kultur und Wissenschaft (`DyNano').
I.\ Latella acknowledges financial support from the Ministerio de Ciencia, Innovación y Universidades of the Spanish Government through Grants No. PID2021-126570NBI00 and PID2024-156516NBI00 (MICIU/FEDER, UE).
\end{acknowledgments}

\onecolumngrid
\vspace{5mm}
\begin{center}
\textbf{End Matter} 
\end{center}
\twocolumngrid

\textit{Correlation function}---We model the particle as an electric dipole with dipole moment $\mathbf{p}(\omega)$ confined within a small volume $V$ at position $\mathbf{r}_p$. The particle radiates into free space in the presence of fluctuating environmental electric and magnetic fields, denoted by $\mathbf{E}^b(\mathbf{r},\omega)$ and $\mathbf{H}^b(\mathbf{r},\omega)$. The dipole moment $\mathbf{p}(\omega)$ consists of two contributions: an intrinsic fluctuating component $\mathbf{p}^f(\omega)$, and an induced component due to the environmental electric field, $\mathbf{p}^\mathrm{ind}(\omega)=\varepsilon_0 \underline{\alpha}\mathbf{E}^b(\mathbf{r}_p,\omega)$. Hence, the total dipole moment is $\mathbf{p}(\omega)=\mathbf{p}^f(\omega)+\mathbf{p}^\mathrm{ind}(\omega)$.
By accounting for radiation from both the particle and the environmental fields, the Fourier components of the electric and magnetic fields $\mathbf{E}(\mathbf{r},t)$ and $\mathbf{H}(\mathbf{r},t)$ at an observation point $\mathbf{r}$ can be approximated as~\cite{Lakhtakia,Chaumet_2009,Abraham_2017}
\begin{equation}
\mathbf{E}(\mathbf{r},\omega)
=
\begin{cases}
\mathbf{E}^b(\mathbf{r},\omega)+ \frac{k_0^2}{\varepsilon_0}\mathds{G}^{\rm EE, vac}(\mathbf{r}, \mathbf{r}_p) \mathbf{p}(\omega)
& \mathbf{r}\notin V \\
\mathbf{E}^b(\mathbf{r},\omega)+ \frac{1}{\varepsilon_0}\left(- \frac{1}{3V} + i\frac{k_0^3 }{6\pi}\right)\mathbf{p}(\omega)
& \mathbf{r}\in V
\end{cases}
\label{local_E_field}
\end{equation}
and
\begin{equation}
\mathbf{H}(\mathbf{r},\omega)
=
\begin{cases}
\mathbf{H}^b(\mathbf{r},\omega)+ \frac{k_0^2}{\varepsilon_0}\mathds{G}^{\rm HE, vac}(\mathbf{r}, \mathbf{r}_p) \mathbf{p}(\omega)
& \mathbf{r}\notin V \\
\mathbf{H}^b(\mathbf{r},\omega)
& \mathbf{r}\in V
\end{cases}.
\label{local_H_field}
\end{equation}

The correlation function in Eq.~(\ref{Eq:correlation_function}) corresponds to the components of the correlation matrix
\begin{equation}
		\langle \mathbf{E} (\mathbf{r}, t)\otimes \mathbf{H}(\mathbf{r}, t) \rangle_{\rm eq}^{\rm sym} = \int_0^\infty \frac{\rd \omega}{2 \pi} 2\Re \langle \mathbf{E}(\mathbf{r},\omega)\otimes \mathbf{H}^*(\mathbf{r},\omega)\rangle,
		\label{Eq:correlation_function_app}
\end{equation}
where the average on right-hand side is computed at thermal equilibrium with symmetrized order for the fields. Furthermore, we define $\mathds{Y}(\mathbf{r};\omega)$ according to
\begin{equation}
2\Re\langle \mathbf{E}(\mathbf{r},\omega)\otimes \mathbf{H}^*(\mathbf{r},\omega)\rangle
=
\frac{2 k_0^3}{\varepsilon_0c}\Theta(\omega,T)
\Im \mathds{Y}(\mathbf{r};\omega).
\end{equation}
The matrix $\mathds{Y}(\mathbf{r};\omega)$ is obtained using Eqs.~(\ref{local_E_field}) and (\ref{local_H_field}) for observation points $\mathbf{r}\notin V$ and evaluating the averages by means of the fluctuation-dissipation theorem, from which~\cite{Manjavacas,Riccardo}
\begin{equation}
\langle \mathbf{p}^f(\omega)\otimes\mathbf{p}^{f*}(\omega)\rangle =\frac{2\varepsilon_0}{\omega}\Theta(\omega,T)\underline{\chi}
\end{equation}
with the susceptibility $\underline{\chi}=(\underline{\alpha} - \underline{\alpha}^{\dagger})/2i  - k_0^3\underline{\alpha}\underline{\alpha}^{\dagger}/6\pi$, while~\cite{GSA,Eckhardt}
\begin{equation}
\langle \mathbf{E}^b(\mathbf{r},\omega)\otimes \mathbf{E}^{b*}(\mathbf{r}',\omega)\rangle
=\frac{2 k_0}{\varepsilon_0c}\Theta(\omega,T)\mathrm{Im}[\mathds{G}^{\rm EE, vac}(\mathbf{r},\mathbf{r}')]
\end{equation}
and
\begin{equation}
\langle \mathbf{E}^b(\mathbf{r},\omega)\otimes \mathbf{H}^{b*}(\mathbf{r}',\omega)\rangle
=i\frac{2 k_0}{\varepsilon_0c}\Theta(\omega,T)\mathrm{Re}[\mathds{G}^{\rm HE, vac}(\mathbf{r},\mathbf{r}')].
\end{equation}
Accordingly, $\mathds{Y}(\mathbf{r};\omega)$ is given by Eq.~(\ref{Eq:matrix_Y}) for $\mathbf{r}\notin V$, which allow us to unambiguously determine the electromagnetic AM in vacuum.

In contrast, the electromagnetic momentum density in matter is subject to different definitions or interpretations~\cite{Griffiths}. The Abraham and Minkowski forms of this density are $\mathbf{g}_A(\mathbf{r},\omega)=\mathbf{E}(\mathbf{r},\omega)\times \mathbf{H}^*(\mathbf{r},\omega)/c^2$ and $\mathbf{g}_M(\mathbf{r},\omega)=\mathbf{D}(\mathbf{r},\omega)\times \mathbf{B}^*(\mathbf{r},\omega)$, respectively, where for our nonmagnetic particle $\mathbf{D}(\mathbf{r},\omega)=\varepsilon_0\underline{\varepsilon}(\omega)\mathbf{E}(\mathbf{r},\omega)$ and $\mathbf{B}(\mathbf{r},\omega)=\mu_0 \mathbf{H}(\mathbf{r},\omega)$. Here, $\underline{\varepsilon}(\omega)$ is the permittivity tensor, see below. 
Since $\mathrm{Re}[\mathds{G}^{\rm HE, vac}(\mathbf{r},\mathbf{r}')]$ vanishes for $\mathbf{r}'=\mathbf{r}$, applying Eqs.~(\ref{local_E_field}) and (\ref{local_H_field}) at an observation point $\mathbf{r}_p=\mathbf{r}\in V$ together with the fluctuation-dissipation theorem yields $\langle\mathbf{g}_A(\mathbf{r}_p,\omega)\rangle=\langle\mathbf{g}_M(\mathbf{r}_p,\omega)\rangle=\mathbf{0}$. Hence, the electromagnetic AM inside the particle can be neglected in the dipole approximation.

\begin{figure}
    \centering
	\includegraphics[width=\columnwidth]{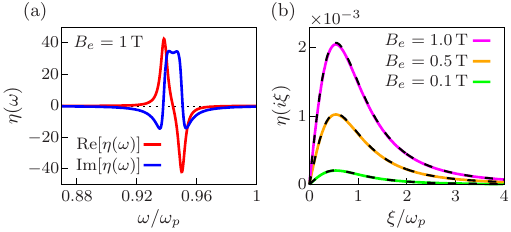}
	\caption{
		\label{Fig:polarizability} Out-of-diagonal component of the polarizability tensor per unit volume of an InSb particle according to Eq.~(\ref{Eq:eta_full_model}). In (a), we show the real and imaginary parts of $\eta(\omega)$ for an external magnetic field of magnitude $B_e=1\,$T, while the function $\eta(\ri\xi)$ is plotted in (b) for different values of the field. The dashed black lines in (b) correspond to the approximation given by Eq.~(\ref{Eq:eta_approx}) in the main text.}
\end{figure}

\textit{Polarizability}---For MO doped semiconductors where the magnetic field points in the $z$-direction, the permittivity tensor has the form~\cite{Palik,Moncada}
\begin{equation}
   \underline{\varepsilon}(\omega) = \begin{pmatrix} \varepsilon_{1} & -\ri\varepsilon_{2} & 0 \\ 
   \ri\varepsilon_{2} & \varepsilon_{1} & 0 \\ 0 & 0 & \varepsilon_{3}	\end{pmatrix}
   \label{Eq:permittivity_tensor}
\end{equation}
with the optical properties described by
\begin{align}
\frac{\varepsilon_{1}}{\varepsilon_\infty}&=1+\frac{\omega_L^2-\omega_T^2}{\omega_T^2-\omega^2-\ri\Gamma\omega}-\frac{\omega_p^2(\omega+\ri\gamma)}{\omega[(\omega+\ri\gamma)^2-\omega_c^2]},\label{Eq:varepsilon_1}\\
\frac{\varepsilon_{2}}{\varepsilon_\infty}&=\frac{\omega_p^2\omega_c}{\omega[(\omega+\ri\gamma)^2-\omega_c^2]}\label{Eq:varepsilon_2},\\
\frac{\varepsilon_{3}}{\varepsilon_\infty}&=1+\frac{\omega_L^2-\omega_T^2}{\omega_T^2-\omega^2-\ri\Gamma\omega}-\frac{\omega_p^2}{\omega(\omega+\ri\gamma)}\label{Eq:varepsilon_3}.
\end{align}
Here, $\omega_L$ and $\omega_T$ are the longitudinal and transverse optical phonon frequencies, $\Gamma$ is the phonon damping constant, and $\gamma$ is the free carrier damping constant. As introduced in the main text, $\varepsilon_\infty$ is the infinite-frequency dielectric constant, $\omega_p=(\frac{n_ce^2}{m^*\varepsilon_0\varepsilon_\infty})^{1/2}$ is the plasma frequency of free carriers with density $n_c$ and effective mass $m^*$, and $\omega_c=eB_e/m^*$ is the cyclotron frequency.
For InSb, we take the parameters~\cite{Palik,Law} $\varepsilon_\infty=15.7$, $\omega_L=3.62\times10^{13}\,$rad/s, $\omega_T=3.39\times10^{13}\,$rad/s, $n_c=1.36\times10^{19}\,$cm$^{-3}$, $m^*=7.29\times10^{-32}\,$kg, $\Gamma=5.65\times10^{11}\,$rad/s, and $\gamma=10^{12}\,$rad/s, so the plasma frequency takes the value $\omega_p=1.86\times10^{14}\,$rad/s.

Furthermore, the polarizability tensor of the particle is given by~\cite{Carminati,Albaladejo} $\underline{\alpha}(\omega)= ( \mathds{1}-\ri k_0^3 \underline{\alpha}_{0}/6\pi )^{-1} \underline{\alpha}_{0}$, with the quasistatic polarizability $\underline{\alpha}_{0}(\omega)=4\pi R_p^3(\underline{\varepsilon}-\mathds{1})(\underline{\varepsilon}+2\mathds{1})^{-1}$ which depends on the permittivity tensor (\ref{Eq:permittivity_tensor}). With simple algebraic manipulations, the out-of-diagonal component $\alpha_{12}$ in Eq.~(\ref{Eq:polarizabilty_single_particle}) is found to be
\begin{equation}
\alpha_{12}(\omega)=-\frac{12\pi R_p^3\,\ri \varepsilon_2}{(\varepsilon_1+2)^2+(\ri \varepsilon_2)^2 + K},
\label{Eq:alpha_12_full}
\end{equation}
where
\begin{equation}
\begin{split}
K&= -\frac{4\ri \omega^3R_p^3}{3c^3}[\varepsilon_1+\varepsilon_1^2+(\ri\varepsilon_2)^2-2]\\
&-\frac{4 \omega^6R_p^6}{9c^6}[(\varepsilon_1-1)^2+(\ri\varepsilon_2)^2] 
\end{split}
\end{equation}
accounts for the radiative correction. From Eq.~(\ref{Eq:alpha_12_full}), we highlight that the polarizability asymptotically behaves as
\begin{equation}
 \omega^3\alpha_{12}(\omega)\sim i\frac{27\pi c^6 \varepsilon_\infty \omega_c \omega_p^2}{ R_p^3(\varepsilon_\infty-1)^2\omega^6}
\label{Eq:asym_rad_correction}
\end{equation}
for $\omega\to\infty$.
Hence, the radiative correction formally ensures that $s(\omega)$ given by Eq.~(\ref{Eq:small_s}) vanishes as $|\omega|\to\infty$ in the first quadrant of the complex $\omega$-plane, as we assumed when computing $\int_\mathcal{C} \rd \omega\, s(\omega)$. The radiative correction can be neglected when evaluating the polarizability at imaginary frequencies.

Dividing $\alpha_{12}(\omega)$ by the volume of the particle yields
\begin{equation}
\eta(\omega)=-\frac{9\ri \varepsilon_2}{(\varepsilon_1+2)^2+(\ri \varepsilon_2)^2 + K}.
\label{Eq:eta_full_model}
\end{equation}
We now look for an approximated expression for $\eta(\ri\xi)$. For a magnetic field $B_e$ small enough, the cyclotron frequency $\omega_c=eB_e/m^*$ is small and, from Eq.~(\ref{Eq:varepsilon_2}), we can approximate $\ri\varepsilon_{2}(\ri\xi)\approx-\varepsilon_\infty\omega_p^2\omega_c/\xi^3$, where we also neglected the damping constant $\gamma$. Thus, for small $\omega_c$ and neglecting the radiative correction, we can write $\eta(\ri\xi)\approx 9\varepsilon_\infty\omega_p^2\omega_c/\xi^3(\varepsilon_1+2)^2$.
Finally, taking $\omega_L\approx\omega_T$ and neglecting $\gamma$ in $\varepsilon_1(\ri\xi)$, to leading order in $\omega_c$, we obtain Eq.~(\ref{Eq:eta_approx}) in the main text.
In Fig.~\ref{Fig:polarizability}, we represent $\eta(\omega)$ and $\eta(\ri \xi)$ according to the full model given by Eq.~(\ref{Eq:eta_full_model}). In Fig.~\ref{Fig:polarizability}(b), the dashed black lines correspond to the approximation for $\eta(\ri\xi)$ given by Eq.~(\ref{Eq:eta_approx}).

\end{document}